\documentclass[a4paper,twocolumn,showpacs,preprintnumbers,amsmath,amssymb]{revtex4}
\usepackage{graphics}
\usepackage[dvips]{graphicx}

\begin{document}

\title{Noise induced Hopf bifurcation}{}
\author{I A Shuda$^1$, S S Borysov$^1$ and A I Olemskoi$^2$}
\address{$^1$Sumy State University, 2, Rimskii-Korsakov St., 40007
Sumy, Ukraine}

\address{$^2$Institute of Applied
Physics, Nat. Acad. Sci. of Ukraine, 58, Petropavlovskaya St.,
40030 Sumy, Ukraine}

\begin{abstract}We consider effect of stochastic sources upon
self-organization process being initiated with creation of the
limit cycle induced by the Hopf bifurcation. General relations
obtained are applied to the stochastic Lorenz system to show that
departure from equilibrium steady state can destroy the limit
cycle in dependence of relation between characteristic scales of
temporal variation of principle variables. Noise induced resonance
related to the limit cycle is found to appear if the fastest
variations displays a principle variable, which is coupled with
two different degrees of freedom or more.
\end{abstract}


\maketitle

\section{Introduction}\label{Sec.1}
Interplay between noise and non-linearity of dynamical systems
\cite{1} is known to arrive at crucial changing in behavior of
systems displaying noise-induced \cite{2,2a} and recurrence
\cite{3a,3} phase transitions, stochastic resonance \cite{4a,4},
noise induced pattern formation \cite{a,b}, and noise induced
transport \cite{c,2a}. The constructive role of noise on dynamical
systems includes hopping between multiple stable attractors
\cite{e,f}, stabilization of the Lorenz attractor near the
threshold of its formation \cite{d,g} and stabilization of
resonance related to the limit cycle near the Hopf bifurcation
\cite{d}. Such type behavior is inherent in systems which involve
discrete entities (for instance, in ecological systems individuals
form population stochastically in accordance with random births
and deaths). Examples of substantial alteration of finite systems
under effect of intrinsic noises gives epidemics
\cite{11}--\cite{13}, predator-prey population dynamics
\cite{5,6}, opinion dynamics \cite{10}, biochemical clocks
\cite{15,16}, genetic networks \cite{14}, cyclic trapping
reactions \cite{9}, etc. Within the phase-plane language, phase
transitions po\-in\-ted out present the simplest case, where a
fixed point appears only. We are interested in studying more
complicated situation, when the system under consideration may
display oscillatory behavior related to the limit cycle appearing
as result of the Hopf bifurcation \cite{17,18}. It has long been
conjectured \cite{19} that in some situations the influence of
noise would be sufficient to produce cyclic behavior \cite{20}.
Recent consideration \cite{21} allows the relation between the
stochastic oscillations in the fixed point phase and the
oscillations in the limit cycle phase to be elucidated. In the
last case, making use of co-moving frame allows fluctuations
transverse and longitudinal with respect to the limit cycle to be
effectively decoupled. It appears while the latter fluctuations
are of a diffusive nature, the former ones follow a stochastic
path. To formulate related model we consider the system with a
finite number of constituents related to components of the state
vector \cite{22}
\begin{equation}
 {\bf n}=N{\bf X}+\sqrt{N}{\bf x}.
 \label{0}
\end{equation}
Characteristically, a deterministic component $\bf X$ is
proportional to total system size $N<\infty$, whereas a random one
${\bf x}$ is the same to its square root. In the limit of infinite
particle numbers $N\to\infty$, such systems are faithfully
described by deterministic equations to find time dependence ${\bf
X}(t)$, which addresses the behavior of the system on a mean-field
level. On the other hand, a systematic study of corrections due to
finite system size can capture the behavior of fluctuations ${\bf
x}(t)$ about the mean-field solution. These fluctuations are
governed with the Langevin equations, however, in difference of
approach \cite{21}, we consider multiplicative noises instead of
additive ones, on the one hand, and nonlinear forces instead of
linear ones, on the other. Within such framework, the aim of the
present paper is to extend analytical descriptions \cite{21} of
finite-size stochastic effects to systems where noises play a
crucial role with respect to periodic limit cycle solution
creation or its supression. We will show that character of the
system behavior is determined with relation between scales of
temporal variation of principle variables and their coupling.

The paper is organized along the following lines. In Section
\ref{Sec.2}, we obtain conditions of the limit cycle creation
making use the pair of stochastic equations with nonlinear forces
and multiplicative noises. Sections \ref{Sec.3}, \ref{Sec.4} are
devoted to consideration of these conditions on the basis of
stochastic Lorenz system with different regimes of principle
variables slaving. According to Section \ref{Sec.3} the limit
cycle is created only in the case if the most fast variation
displays a principle variable, which is coupled nonlinearly with
two degrees of freedom or more. Opposite case is studied in
Section \ref{Sec.4} to show that the limit cycle disappears in non
equilibrium steady state. Section \ref{Sec.5} concludes our
consideration.

\section{Statistical picture of limit cycle}\label{Sec.2}

According to the theorem of central manifold \cite{17}, to
achie\-ve a closed description of a limit cycle it is enough to
use only two variables $x_\alpha$, $\alpha=1,2$. In such a case,
stochastic evolution of the system under investigation is defined
by the Langevin equations \cite{23}
\begin{equation}
 \dot{x}_\alpha=f^{(\alpha)}+\mathcal{G}_{\alpha}\zeta(t),\quad \alpha=1,2
 \label{1}
\end{equation}
with forces $f^{(\alpha)}=f^{(\alpha)}(x_1,x_2)$ and noise
amplitudes $\mathcal{G}_{\alpha}=\mathcal{G}_{\alpha}(x_1,x_2)$,
being functions of stochastic variables $x_\alpha$, $\alpha=1,2$,
and white noise $\zeta(t)$ determined as usually:
$\langle\zeta(t)\rangle=0$,
$\langle\zeta(t)\zeta(t')\rangle=\delta(t-t')$. Our principle
assumptions are as follows: (i) white noise $\zeta(t)$ is equal
for both degrees of freedom $x_\alpha$; (ii) microscopic transfer
rates are non correlated for different variables $x_\alpha$. Then,
the probability distribution function
$\mathcal{P}=\mathcal{P}(x_1,x_2;t)$ is determined by the
Fokker-Planck equation
\begin{equation}
\frac{\partial\mathcal{P}}{\partial t}+\frac{\partial
J^\alpha}{\partial x_\alpha}=0,
 \label{3}
\end{equation}
where sum over repeated Greek indexes $\alpha=1,2$ is meant and
components of the probability current take the form
\begin{equation}
J^{(\alpha)}\equiv
\mathcal{F}^{(\alpha)}\mathcal{P}-\frac{1}{2}\frac{\partial
}{\partial
x_\beta}\left(\mathcal{G}_\alpha\mathcal{G}_\beta\mathcal{P}\right),
 \label{4}
\end{equation}
with the generalized forces
\begin{equation}
\mathcal{F}^{(\alpha)}=f^{(\alpha)}+
\lambda\frac{\partial\left(\mathcal{G}_\alpha\mathcal{G}_\beta
\right)}{\partial x_\beta} \label{4a}
\end{equation}
being determined with choice of the calculus parameter $\lambda\in
[0,1]$. Within the steady state, the components of the probability
current take constant values $J^{(\alpha)}_0$ and the system
behaviour is defined by the following equations:
\begin{eqnarray}
 \frac{\partial}{\partial
x_1}\left(\mathcal{G}_1^2\mathcal{P}\right)+\frac{\partial
}{\partial
x_2}\left(\mathcal{G}_1\mathcal{G}_2\mathcal{P}\right)-2\mathcal{F}^{(1)}\mathcal{P}=-2J^{(1)}_0,\\
 \frac{\partial }{\partial
x_1}\left(\mathcal{G}_1\mathcal{G}_2\mathcal{P}\right)+\frac{\partial
}{\partial
x_2}\left(\mathcal{G}_2^2\mathcal{P}\right)-2\mathcal{F}^{(2)}\mathcal{P}=-2J^{(2)}_0.
 \label{6}
\end{eqnarray}

Multiplying the first of these equations by factor $\mathcal{G}_2$
and the second one by $\mathcal{G}_1$ and then subtracting
results, we arrive at the explicit form of the probability
distribution function as follows:
\begin{eqnarray} \label{7}
\mathcal{P}\left(x_1,x_2\right)=\frac{J^{(1)}_0\mathcal{G}_2\left(x_1,x_2\right)
-J^{(2)}_0\mathcal{G}_1\left(x_1,x_2\right)}{\mathcal{D}\left(x_1,x_2\right)},\nonumber\\
\mathcal{D}\left(x_1,x_2\right)\equiv\left(\mathcal{G}_2\mathcal{F}^{(1)}
-\mathcal{G}_1\mathcal{F}^{(2)}\right)\\
+\frac{1}{2}\left[\left(\mathcal{G}_1^2\frac{\partial\mathcal{G}_2}{\partial
x_1}-\mathcal{G}_2^2\frac{\partial\mathcal{G}_1}{\partial
x_2}\right)-\mathcal{G}_1\mathcal{G}_2\left(\frac{\partial\mathcal{G}_1}{\partial
x_1}-\frac{\partial\mathcal{G}_2}{\partial
x_2}\right)\right].\nonumber
\end{eqnarray}
This function diverges at condition
\begin{eqnarray}
2\left(\mathcal{G}_1\mathcal{F}^{(2)}-\mathcal{G}_2\mathcal{F}^{(1)}\right)\nonumber\\
=\left(\mathcal{G}_1^2\frac{\partial\mathcal{G}_2}{\partial
x_1}-\mathcal{G}_2^2\frac{\partial\mathcal{G}_1}{\partial
x_2}\right)-\mathcal{G}_1\mathcal{G}_2\left(\frac{\partial\mathcal{G}_1}{\partial
x_1}-\frac{\partial\mathcal{G}_2}{\partial x_2}\right).
 \label{8}
\end{eqnarray}
that physically means appearance of a domain of forbidden values
of stochastic variables $x_\alpha$, which is bonded with a closed
line of the limit cycle. Characteristically, such a line appears
if the denominator $\mathcal{D}(x_1,x_2)$ of fraction (\ref{7})
includes even powers of both variables $x_1$ and $x_2$.

It is worth to note the analytical expression (\ref{7}) of the
probability distribution function becomes possible due to the
special form of the probability current (\ref{4}), where effective
diffusion coefficient takes the multiplicative form
$\mathcal{D}_{\alpha\beta}=\mathcal{G}_\alpha\mathcal{G}_\beta$.
In general case, this coefficient is known \cite{24} to be defined
with the expression
\begin{equation}
\mathcal{D}_{\alpha\beta}=\sum\limits_{ab}I_{ab}g_\alpha^ag_\beta^b,
 \label{8a}
\end{equation}
where kernel $I_{ab}$ determines transfer rate between microscopic
states $a$ and $b$, whereas factors $g_\alpha^a$ and $g_\beta^b$
are specific noise amplitudes of values $x_\alpha$ related to
these states. We have considered above the simplest case, when the
transfer rate $I_{ab}=I$ is constant for all microscopic states.
As a result, the diffusion coefficient (\ref{8a}) takes the needed
form
$\mathcal{D}_{\alpha\beta}=\mathcal{G}_\alpha\mathcal{G}_\beta$
with cumulative noise amplitudes $\mathcal{G}_\alpha
\equiv\sqrt{I}\sum_{a}g_\alpha^a$ and
$\mathcal{G}_\beta\equiv\sqrt{I}\sum_{b}g_\beta^b$.

\section{Noise induced resonance within Lorenz system}\label{Sec.3}

As the simplest and most popular example of the self-organization
induced by the Hopf bifurcation, we consider modulation regime of
spontaneous laser radiation, whose behaviour is presented in terms
of the radiation strength $E$, the matter polarization $P$ and the
difference of level populations $S$ \cite{22}. With accounting for
stochastic sources related, the self-organization process of this
system is described by the Lorenz equations
\begin{eqnarray}\label{lor}
 \tau_E\dot{E}=[-E+a_E P-\varphi(E)]+g_E\zeta(t),\nonumber\\
 \tau_P\dot{P}=(-P+a_P ES)+g_P\zeta(t),\\
 \tau_S\dot{S}=[(S_e-S)-a_S EP]+g_S\zeta(t).\nonumber
\end{eqnarray}
Here, overdot denotes differentiation over time $t$;
$\tau_{E,P,S}$ and $a_{E,P,S}>0$ are time scales and feedback
constants of related variables, respectively; $g_{E,P,S}$ are
corresponding noise amplitudes, and $S_e$ is driven force. In the
absence of noises $(g_{E}=g_{P}=g_{S}=0)$ and at relation
$\tau_P,\tau_S\ll\tau_E$ between time scales, the system addresses
to limit cycle only in the presence of the nonlinear force
\cite{25}
\begin{equation}
\varphi(E)=\frac{\kappa E}{1+E^2/E_n^2} \label{11ab}
\end{equation}
characterized with parameters $\kappa>0$ and $E_n$. In this
Section, we consider noise effect in the case of opposite relation
$\tau_E\ll\tau_P,\tau_S$ of time scales, when periodic variation
of stochastic variables becomes possible even at suppression of
the force (\ref{11ab}).

It is convenient further to pass to dimensionless variables $t$,
$\zeta$, $E$, $P$, $S$, $g_E$, $g_P$, $g_S$ with making use of the
related scales:
\begin{eqnarray} \label{scale}
\tau_P;\quad \zeta_s=\tau_E^{-1/2};\quad E_{s}=(a_Pa_S)^{-1/2},\nonumber\\
P_s=(a^2_E a_P a_S)^{-1/2},\quad S_s=(a_E a_P)^{-1};\\
g_E^{s}=(\tau_P/a_Pa_S)^{1/2},\quad g_P^s=(\tau_P/a^2_E a_P
a_S)^{1/2},\nonumber\\ g_S^s=\tau_P^{1/2}/a_E a_P.\nonumber
\end{eqnarray}
Then, equations (\ref{lor}) (these equations are reduced to
initial form \cite{22a} if one set there
$X\equiv\sqrt{\sigma/\varepsilon}E$,
$Y\equiv\sqrt{\sigma/\varepsilon}P$, $Z\equiv S_e-S$, $r\equiv
S_e$, and $b\equiv\sigma/\varepsilon$.) take the simple
form
\begin{eqnarray}\label{lor11a}
 \sigma^{-1}\dot{E}=-E+P-\varphi(E)+g_E\zeta(t),\nonumber\\
 \dot{P}=-P+ES+g_P\zeta(t),\\
 (\varepsilon/\sigma)\dot{S}=(S_e-S)-EP+g_S\zeta(t)\nonumber,
\end{eqnarray}
where the time scale ratios
\begin{equation}
\sigma\equiv\tau_P/\tau_E,\quad
 \varepsilon\equiv\tau_S/\tau_E
 \label{time}
\end{equation}
are introduced. In the absence of noises, the Lorenz system
(\ref{lor11a}) is known to show the usual bifurcation in the point
$S_e=1$ and the Hopf bifurcation at the driven force \cite{22a,22}
\begin{equation}
S_e=\frac{\tau_P}{\tau_E}~\frac{\tau_E^{-1}+\tau_S^{-1}+3\tau_P^{-1}}
{\tau_E^{-1}-\tau_S^{-1}-\tau_P^{-1}}.
 \label{cond}
\end{equation}
However, at $g_E=g_P=g_S=0$ the limit cycle is unstable and the
Hopf bifurcation arrives at the strange attractor only.

With switching on noises, the condition $\tau_E\ll\tau_P$ allows
for to put l.h.s. of the first equation (\ref{lor11a}) to be equal
zero. Then, the strength is expressed with the equality
\begin{equation}
E=P+g_E\zeta(t),
 \label{11a}
\end{equation}
whose insertion into the system (\ref{lor11a}) reduces it into
two-dimensional form
\begin{eqnarray}\label{22}
 \dot{P}=-P(1-S)+G_P\zeta(t),\nonumber\\
 \dot{S}=(\sigma/\varepsilon)\left[(S_e-S)-P^2\right]+G_S\zeta(t)
 \end{eqnarray}
with the effective amplitudes of multiplicative noises
\begin{equation}\label{11abc}
\mathcal{G}_P\equiv\sqrt{g_P^2+g_E^2S^2},\quad
\mathcal{G}_S\equiv(\tau_{P}/\tau_{S})\sqrt{g_S^2+g_E^2P^2}
\end{equation}
and the generalized forces
\begin{eqnarray}
 \mathcal{F}^{(P)}=-P(1-S)+\lambda\frac{g_E^2}{\tau_{S}/\tau_{P}}S
 \sqrt{\frac{(g_S/g_E)^2+P^2}{(g_P/g_E)^2+S^2}},\nonumber\\
 \mathcal{F}^{(S)}=(\tau_{P}/\tau_{S})\left[(S_e-S)-P^2\right]
 \\+\lambda\frac{g_E^2}{\tau_{S}/\tau_{P}}P
 \sqrt{\frac{(g_P/g_E)^2+S^2}{(g_S/g_E)^2+P^2}}.
 \nonumber
\end{eqnarray}
In this way, the probability density (\ref{7}) takes infinite
values at condition
\begin{eqnarray} \label{111}
\left(\frac{g_S^2}{g_E^2}+P^2\right)\sqrt{\frac{g_P^2}{g_E^2}+S^2}P(1-S)\nonumber\\+
\left(\frac{g_P^2}{g_E^2}+S^2\right)\sqrt{\frac{g_S^2}{g_E^2}+P^2}\left[(S_e-S)-P^2\right]\\+
\frac{g_E^2}{2}\frac{\sigma}{\varepsilon}\left(\frac{g_S^2}{g_E^2}
+P^2\right)^{\frac{3}{2}}S-\frac{g_E^2}{2}
\left(\frac{g_P^2}{g_E^2}+S^2\right)^{\frac{3}{2}}P=0,\nonumber
\end{eqnarray}
where we choose the simplest case of Ito calculus $(\lambda=0)$.

\begin{figure}
\centering
\includegraphics[width=0.9\columnwidth]{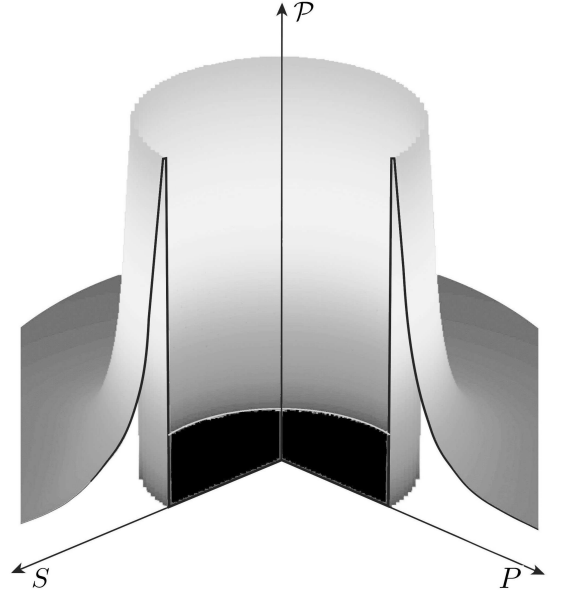}
\caption{Steady-state distribution function (\ref{7}) at
$J^{(P)}_0=1$, $J^{(S)}_0=10$, $\tau_P=\tau_S$, $S_e=0.5$,
$g_E=0.5$, $g_P=1.376$, $g_S=2.5$}\label{prob1.eps}
\end{figure}

\begin{figure}
\centering a\hspace{0.5\columnwidth}b\\
\includegraphics[width=0.48\columnwidth]{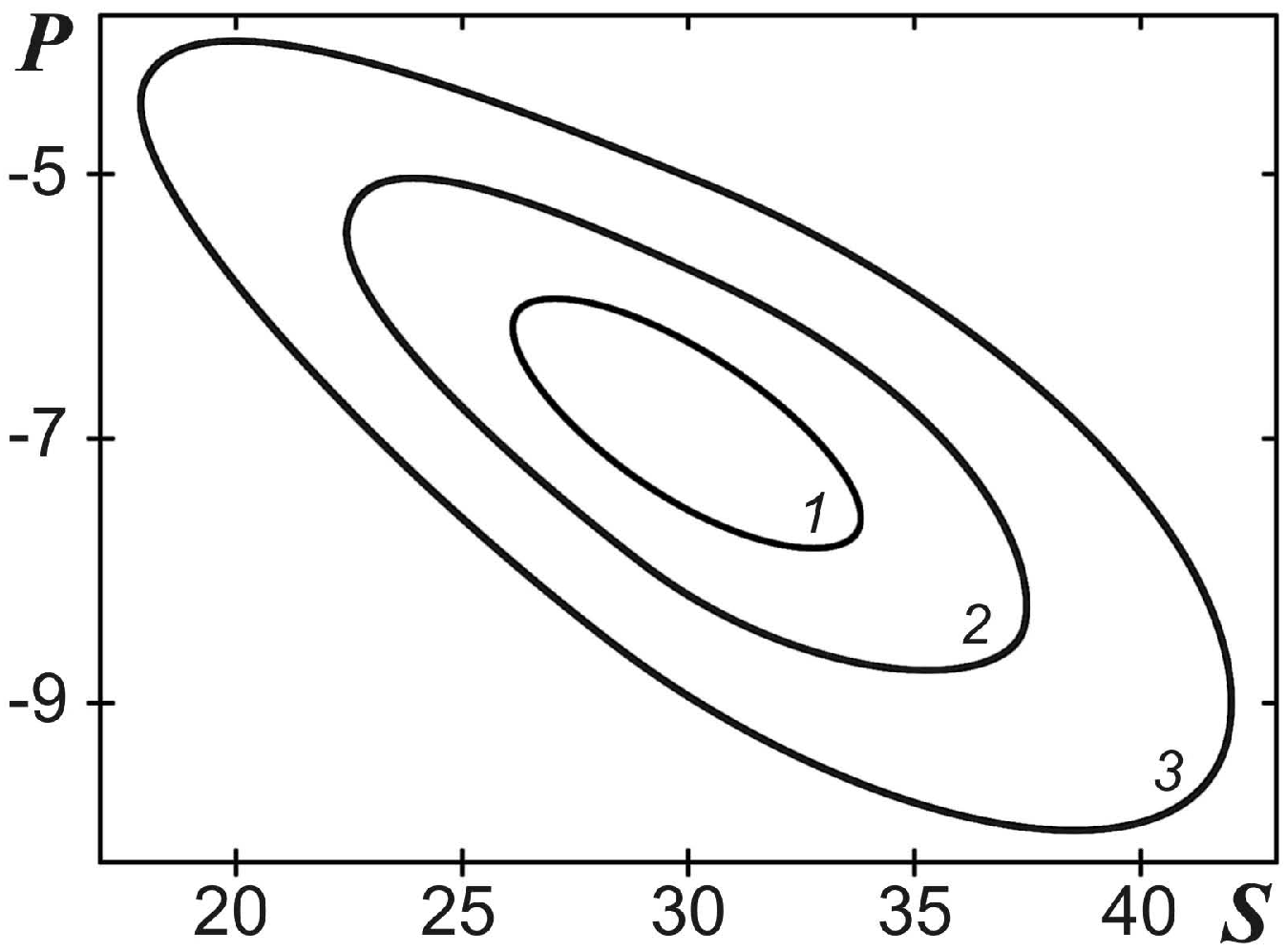}
\includegraphics[width=0.48\columnwidth]{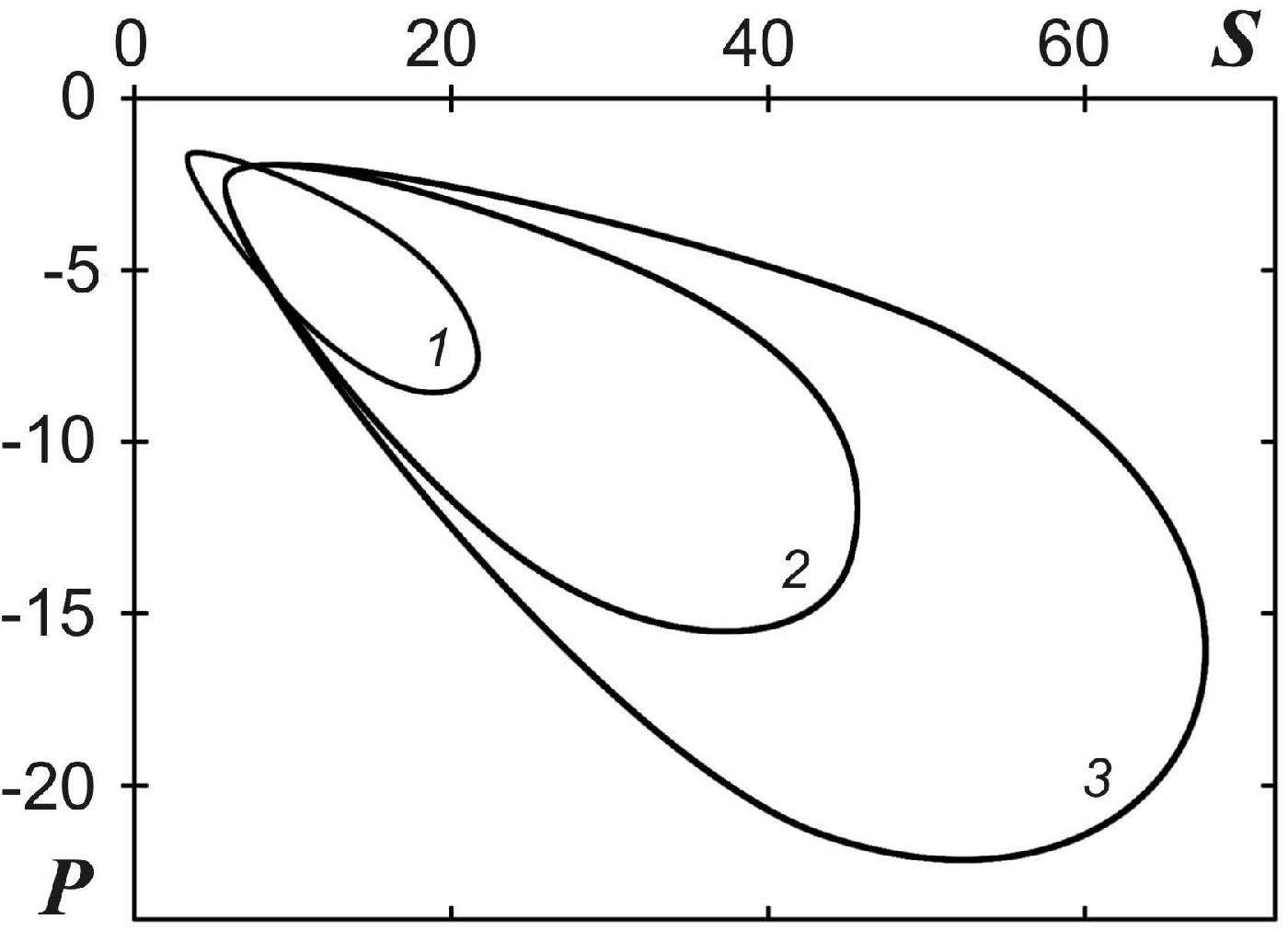}
\caption{Form of limit cycle at: a) $g_E=0.5$, $g_P=11$, $g_S=6$
(curves 1-3 relate to $S_E=0.5, 1.0, 2.0$, correspondingly);
\break b) $S_e=0.5$, $g_P=7.5$, $g_S=6.5$ (curves 1-3 relate to
$g_E=1.0, 0.6, 0.5$, correspondingly)}\label{ab.eps}
\end{figure}

\begin{figure}
\centering a\hspace{0.5\columnwidth}b\\
\includegraphics[width=0.48\columnwidth]{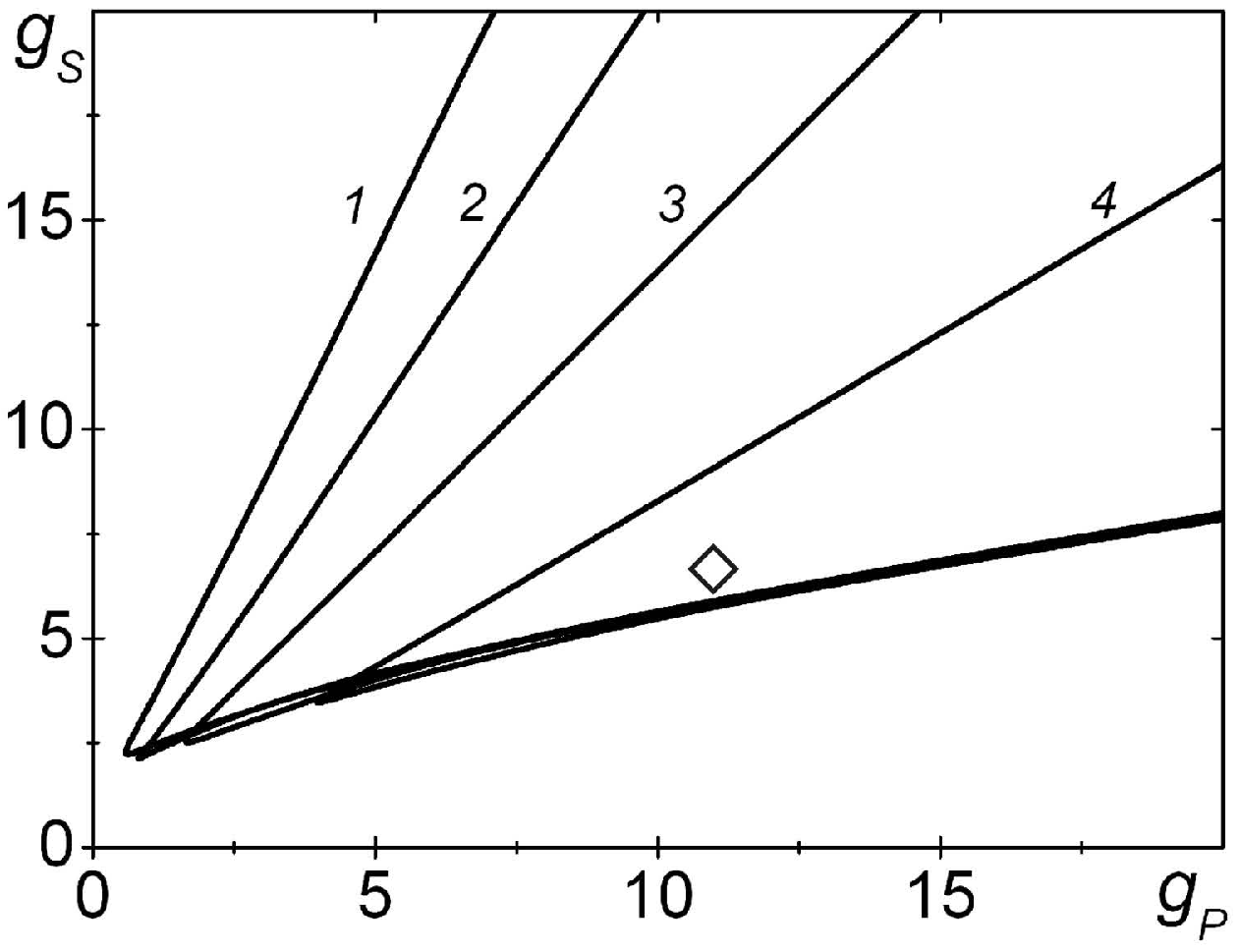}
\includegraphics[width=0.48\columnwidth]{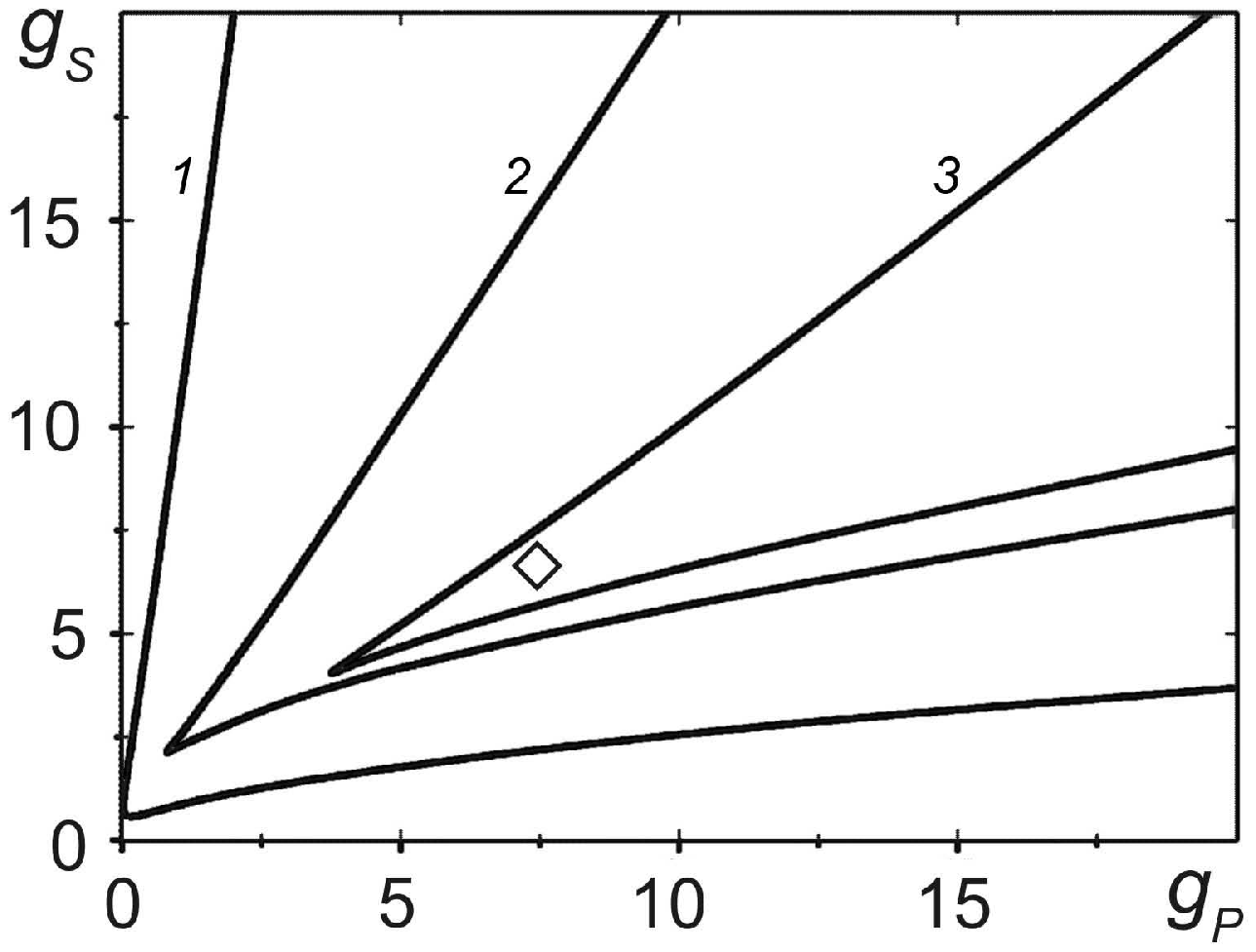}
\caption{Phase diagrams of the limit cycle creation determined
with Eq.(\ref{111}) at: a) $g_E=0.5$, curves 1-4 correspond to
$S_e=0.0, 0.5, 1.0, 2.0$, respectively; b) $S_e=0.5$, curves 1-3
correspond to $g_E=0.1, 0.5, 1.0$, respectively (diamonds relate
to the values $g_P$ and $g_S$, for which limit cycles in
Fig.\ref{ab.eps} are depicted)}\label{se.eps}
\end{figure}

Reduced Lorenz system (\ref{22}) has two-dimensional form
(\ref{1}), where the role of variables $x_1$ and $x_2$ play the
matter polarization $P$ and the difference of level populations
$S$. According to the distribution function (\ref{7}) shown in
Fig.\ref{prob1.eps}, the stochastic
variables $P$ and $S$ are realized with non-zero probabilities out
off the limit cycle only, whereas in its interior the domain of
forbidden values $P$, $S$ appears. That is principle difference
from the deterministic limit cycle, which bounds a domain of
unstable values of related variables. The form of this domain is
shown in Fig.\ref{ab.eps}
at different values of the noise amplitudes $g_E$, $g_P$, $g_S$
and driven force $S_e$. It is seen, this domain grows with
increase of the driven force $S_E$, whereas increase of the force
fluctuations $g_E$ shrinks it. On the other hand, phase diagrams
depicted in Fig.\ref{se.eps} show
the strengthening noises of both polarization and difference of
level populations enlarges domain of the limit cycle creation (in
this way, force noise $g_E$ shrinks this domain from both above
and below, whereas increase of driven force $S_e$ makes the same
from above only).

\section{Lorenz system without limit cycle}\label{Sec.4}

At condition $\tau_{P}\ll\tau_{E},\tau_{S}$, the deterministic
system \break $(g_{E,P,S}=0)$ has a limit cycle only at large
intensity $\kappa$ of non linear force (\ref{11ab}) \cite{25}. In
this case, it is convenient to measure the time $t$ in the scale
$\tau_E$ and replace $\tau_P$ by $\tau_E$ in set of scales
(\ref{scale}). Then, one obtains the relation (cf. Eq.(\ref{11a}))
\begin{equation}
P=ES+g_P\zeta(t)
\end{equation}
and the Lorenz system (\ref{lor11a}) is reduced to two-dimensional
form
\begin{eqnarray}
 \dot{ E}=-\left[E(1-S)+\varphi(E)\right]+\mathcal{G}_E\zeta(t),\nonumber\\
 \dot{S}=\varepsilon^{-1}\left[S_e-S(1+ E^2)\right]+\mathcal{G}_S\zeta(t)\label{2b}
\end{eqnarray}
with the effective noise amplitudes
\begin{eqnarray}
\mathcal{G}_E\equiv\sqrt{g_P^2+g_E^2},\quad
\mathcal{G}_S\equiv\varepsilon^{-1}\sqrt{g_S^2+g_P^2 E^2}.
 \label{3b}
\end{eqnarray}
The generalized forces are as follows:
\begin{eqnarray}\label{5b}
 \mathcal{F}^{(E)}=-\left[E(1-S)+\varphi(E)\right],\nonumber\\
 \mathcal{F}^{(S)}=\varepsilon^{-1}\left[(S_e-S)-SE^2\right]
 \\+\lambda\frac{g_P^2}{\varepsilon}E
 \sqrt{\frac{1+(g_E/g_P)^2}{(g_S/g_P)^2+E^2}}.\nonumber
\end{eqnarray}
The probability distribution function (\ref{7}) diverges at
condition
\begin{eqnarray} \label{A}
\frac{\left(g_S/g_P\right)^2+E^2}{1+\left(g_E/g_P\right)^2}\left[\varphi(E)+E(1-S)\right]\nonumber\\+
\sqrt{\frac{\left(g_S/g_P\right)^2+E^2}{1+\left(g_E/g_P\right)^2}}\left[S_e-S(1+E^2)\right]
\\+\left(\lambda-\frac{1}{2}\right)g_P^2E=0,\nonumber
\end{eqnarray}
being the equation, which does not include even powers of the
variable $S$.

\begin{figure}
\includegraphics[width=0.95\columnwidth]{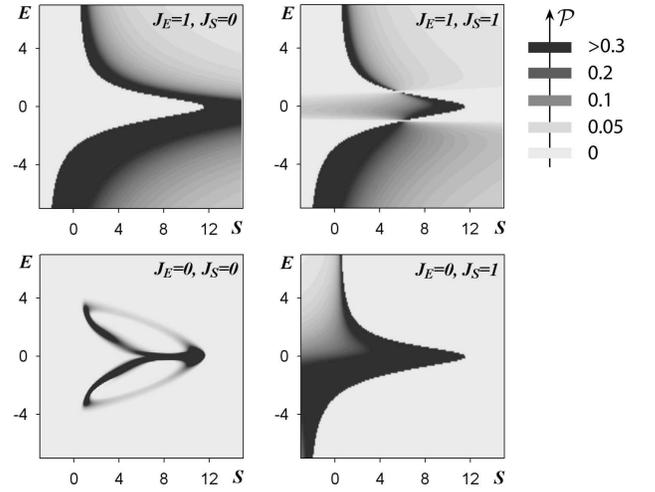}
\caption{Steady-state probability distribution as function of the
radiation strength $E$ and the difference of level populations $S$
at conditions $\tau_{P}\ll\tau_{E}=\tau_{S}$ for parameters
$\kappa=10$, $S_e=11.6$, $g_E=0.2$, $g_P=0.2$, $g_S=0.2$ and
different probability currents $J_0^{(E)}$, $J_0^{(S)}$ (shown on
the related panels)}\label{prob.eps}
\end{figure}

As a result, one can conclude that departure from equilibrium
steady state destroys deterministic limit cycle at the relation
$\tau_{P}\ll\tau_{E},\tau_{S}$ between characteristic scales. This
conclusion is confirmed with Fig.\ref{prob.eps} that shows
divergence of the
probability distribution function on the limit cycle of variation
of the radiation strength $E$ and the difference of level
populations $S$ at zeros probability currents $J_0^{(E)}$ and
$J_0^{(S)}$ only. With increase of these currents the system
escapes from equilibrium steady state and maximum of the
distribution function shifts to non-closed curves being determined
with equation (\ref{A}).

\section{Conclusion}\label{Sec.5}

We have considered effect of stochastic sources upon
self-organization process being initiated with creation of the
limit cycle induced by the Hopf bifurcation. In Sections
\ref{Sec.3}--\ref{Sec.4}, we have applied general relations
obtained in Section \ref{Sec.2} to the stochastic Lorenz system to
show that departure from equilibrium steady state can destroy or
create the limit cycle in dependence of relation between
characteristic scales of temporal variation of principle
variables.

Investigation of the Lorenz system with different re\-gi\-mes of
principle variables slaving shows that additive noises can take
multiplicative character if one of these noises has many fewer
time scale than others. In such a case, the limit cycle may be
created if the most fast variable is coupled with more than two
slow ones. However, the case considered in Section \ref{Sec.4}
shows such dependence is not necessarily to arrive at limit cycle,
as within adiabatic condition $\tau_P\ll\tau_E$, both noise
amplitude $G_S(E)$ and generalized force $\mathcal{F}^{(S)}(E)$,
determined with Eqs. (\ref{3b}), (\ref{5b}), are functions of the
squared strength $E^2$ only. The limit cycle is created if the
fastest variations displays a principle variable, which is coupled
with two different degrees of freedom or more. Indeed, at the
relation $\tau_E\ll\tau_P,\tau_S$ of relaxation times considered
in Section \ref{Sec.3}, variations of the strength $E$ in
nonlinear terms of two last equations (\ref{lor11a}) arrive at
double-valued dependencies of the noise amplitudes $G_P$ and $G_S$
on both difference of level populations and polarization, which
are appeared in Eqs.(\ref{11abc}) as squares $S^2$ and $P^2$. This
appears physically as noise induced resonance related to the limit
cycle created by the Hopf bifurcation, that has been observed both
numerically \cite{d} and analytically \cite{26}.

\end{document}